\documentclass[onecolumn,prd,epsf]{revtex4}
\usepackage{graphicx}
\usepackage{dcolumn}
\usepackage{bm}
\begin{document}
\tolerance=100000

\def\dis{\displaystyle}
\def\beq{\begin{equation}}
\def\eeq{\end{equation}}
\def\barr{\begin{array}}
\def\earr{\end{array}}

\def\bothpart{\mathrel{\rlap{\raise1.6ex\hbox{$\leftrightarrow$}}
                    {\lower.0ex\hbox{$\partial$}}}}

\def\Ecm{\ifmmode{E_{\mathrm{cm}}}\else{$E_{\mathrm{cm}}$}\fi}
\def\lsim{\buildrel{\scriptscriptstyle <}\over{\scriptscriptstyle\sim}}
\def\gsim{\buildrel{\scriptscriptstyle >}\over{\scriptscriptstyle\sim}}
\def \lum{{\cal L}}

\def\lapp{\mathrel{\rlap{\raise.5ex\hbox{$<$}}
                    {\lower.5ex\hbox{$\sim$}}}}
\def\gapp{\mathrel{\rlap{\raise.5ex\hbox{$>$}}
                    {\lower.5ex\hbox{$\sim$}}}}

\title{Top off the unparticle}
\author{Debajyoti Choudhury and Dilip Kumar Ghosh}
 \affiliation{Department of Physics and Astrophysics, University of Delhi, Delhi 110 007, India.}
\begin{abstract}
{\noindent\normalsize 
The existence of an exactly scale invariant sector 
possessing a non-trivial infrared fixed point at a
higher energy scale and its possible communication 
with the Standard Model particles through a heavy 
messenger sector has been shown to lead to curious
unparticle effects. We demonstrate that top physics 
at the Tevatron can already constrain such theories. 
We also consider possible improvements 
at the LHC and delineate some striking signatures.
}
\end{abstract}
\maketitle

The top quark, with a mass very close to the electroweak symmetry
breaking scale, plays a unique role in understanding the Standard
Model (SM).  For example, the agreement between the directly measured
value $m_t$, and the one indicated by precision
measurements~\cite{lepwwg}, has played a crucial role
in testing the SM to the loop level.  Similarly, a study of the
production and the properties of the top quark at the TeV colliders
can be used as a `low' energy probe for any (`high scale') new physics
beyond the SM~\cite{Hill:1993hs}.  At the Tevatron, this has already
led to very fruitful
investigations~\cite{Cabrera:2006ya,Lannon:2006vm} and one expects a
top factory such as the LHC to provide a very productive arena for
studying the SM as well as the beyond the SM 
physics~\cite{Beneke:2000hk}. 
Recent discussions have addressed the possibility 
of identifying Kaluza Klein (KK) gluons of the bulk Randall-Sundrum Model 
through their r\^ole in 
$t \bar t$ production at the Tevatron or the LHC~\cite{KKgluon};
and, similarly, of probing models with extended color sectors~\cite{axigluon}.
The use of spin-spin correlations have also been advocated to increase 
the sensitivity, whether for $Z'$ searches~\cite{Zpr}, or for 
KK excitations of the graviton~\cite{Arai:2004yd}.
 
While all of the above are but examples of new hypothetical particles
playing a detectable r\^ole in $t\bar t$ production, note that the
latter could also be affected even in the absence of a relatively
light new particle. A striking example is afforded by the recently
introduced ``unparticle''~\cite{Georgi:2007ek}, a consequence of having
a scale invariant sector with a non-trivial infrared fixed point. As
is well known, an exact scale invariance requires that the mass
spectrum be either continuous or that all masses be zero. Thus, the SM,
with its discrete mass spectrum manifestly breaks scale
invariance. However, this does not preclude the existence of a new
sector that is so weakly coupled to the SM that we have been unable to
probe it experimentally. If this new physics were to be described by a
nontrivial scale invariant theory sector with an infrared fixed point
(examples being afforded by a vector-like non-abelian gauge theory
with a large number of massless fermions as studied by Banks and Zaks
(BZ)~\cite{Banks:1981nn}, or certain nonlinear sigma
models~\cite{Braaten:1985is}), it would manifest itself in the
existence of asymptotic states that are not particle-like but are
``unparticles''~\cite{Georgi:2007ek} in the sense of having a continuous
mass spectrum. (It has been demonstrated~\cite{Stephanov} that
the unparticle can be deconstructed as the limiting case of an
infinite tower of particles of different masses with a regular mass
spacing.) With the interaction of the two sectors being mediated by
an unspecified superheavy messenger sector, at low energies, it can be
parametrized in terms of effective Lagrangians. Curiously, the
unparticle operators $O_{\cal U}$ (which can have any possible spin
structure) need not have an integral mass dimension. Rather, the
final state spectrum corresponding to an operator of dimension
$d_{\cal U}$ (possibly fractional) resembles that of $d_{\cal U}$
massless particles. This aspect is also reflected in the 
structure of the unparticle propagator\cite{Georgi:2007ek,
Georgi:2007si, Cheung:2007ue}. Not surprisingly, these novel features 
lead to curious phenomenological consequences~\cite{Georgi:2007ek,
Georgi:2007si, Cheung:2007ue, fcnc_unp, other_unp, 
Stephanov, Choudhury:2007js}

In this note, we examine the possible consequences of such a sector on
top physics, in particular the constraints that the current
measurements on $t \bar t$ production at the
Tevatron~\cite{Cabrera:2006ya,Lannon:2006vm} imply for such theories.
The relevant operators in the effective Lagrangian are given by 
\beq
\barr{rcl} {\cal L} & = & 
\dis \Lambda^{-d_{\cal U}} {\cal O}_S \left[ -c_t 
\bar t  \not\bothpart \gamma_5  t  \;  
+ G_{\mu \nu} \!\left(\frac{c_1}{4}  G^{\mu \nu} + \frac{c_2}{4}
\widetilde G^{\mu \nu} \right) \right] 
\\[2ex] 
& + & \dis 
\Lambda^{1 -d_{\cal U}}\; 
\sum_q \bar q  \gamma_\eta  (\tilde v^q + \tilde a^q  \gamma_5)  q \; 
{\cal O}^\eta_V 
\\[2ex] 
& + & \dis
\Lambda^{-d_{\cal U}}\; {\cal O}_T^{\mu \nu} \; 
\Bigg[ \frac{-1}{4} \;
\sum_q \bar q  \gamma_{\{\mu} \bothpart_{\nu\}}  
              (a_q + b_q  \gamma_5)  q 
\\[2ex]
& & \dis \hspace*{2em}
+ G^{\alpha}_\nu \; \left( a_g \; G_{\mu \alpha} 
                          + b_g \; \widetilde G_{\mu \alpha} 
\right) \Bigg] 
\earr
     \label{lagr} 
\eeq 
where ${\cal O}_{S, V, T}$ respectively denote
scalar, transverse vector and transverse symmetric tensor operator
with mass dimension $d_{\cal U}$ in each case and $\Lambda$ denotes
the characteristic scale of the interaction. 
In the spirit of effective theories, we shall
consider the coefficients to be either unity or zero. And, unless 
stated otherwise, we restrict ourselves to right-handed fermion
fields alone. Note that the coupling of ${\cal O}_S$ to light fermions 
vanishes with the fermion mass.
 Armed with the above, and using the propagators as derived in
Refs.~\cite{Georgi:2007si,Cheung:2007ue}, we may now calculate the
parton-level cross sections for $q \bar q \to t \bar t$ and $g g \to t
\bar t$.  For all our
computations we use the CTEQ-6L1 parton
distributions~\cite{Pumplin:2002vw}, with a choice of $Q^2 = m_t^2$
for the factorization scale. The QCD correction, within the SM,
has been calculated
in Refs.~\cite{nnlo}. In the absence of such 
calculations in generic theories, we use the SM  
$K$-factor for the entire process. In view of its relative smallness, 
any error on this account is expected to be small.

At the Tevatron, the $q\bar q$-initiated process 
dominates the $gg$-initiated one by a factor of over 15, mainly on account 
of the relative sizes of the fluxes. Thus, we expect that even for the 
unparticle-mediated contributions, a similar hierarchy would hold and 
this is borne out by explicit calculations. While several 
unparticle operator can contribute to $t \bar t$ production,
we choose one set of operators (or, equivalently, one new 
partonic process) at a time and
study its effects. 

\begin{figure}[!h]
\begin{center}
\vspace*{1cm}
\includegraphics[width= 9 cm, height= 6 cm]{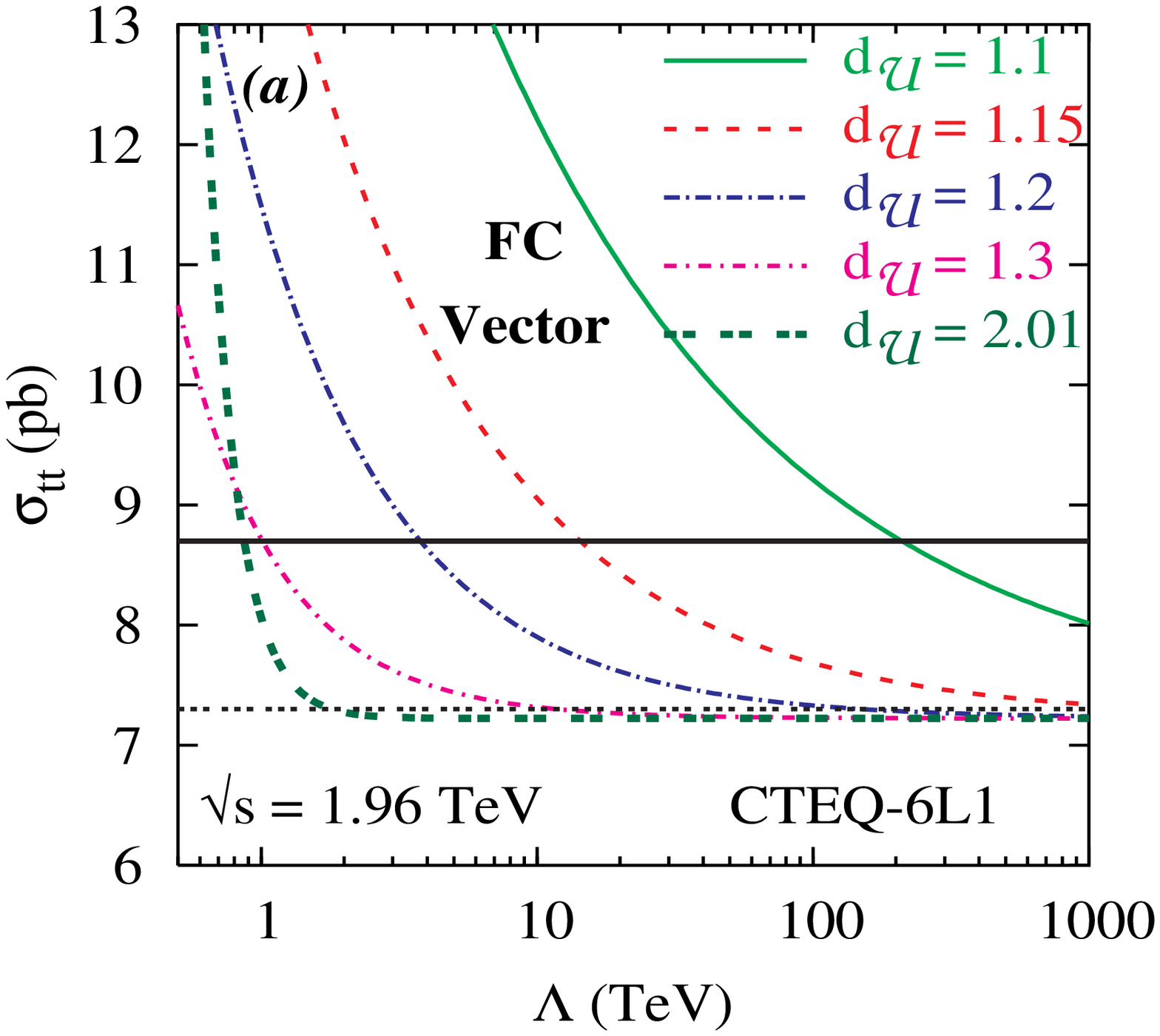}
\vspace*{-2.2cm}
\includegraphics[width= 9 cm, height= 6 cm]{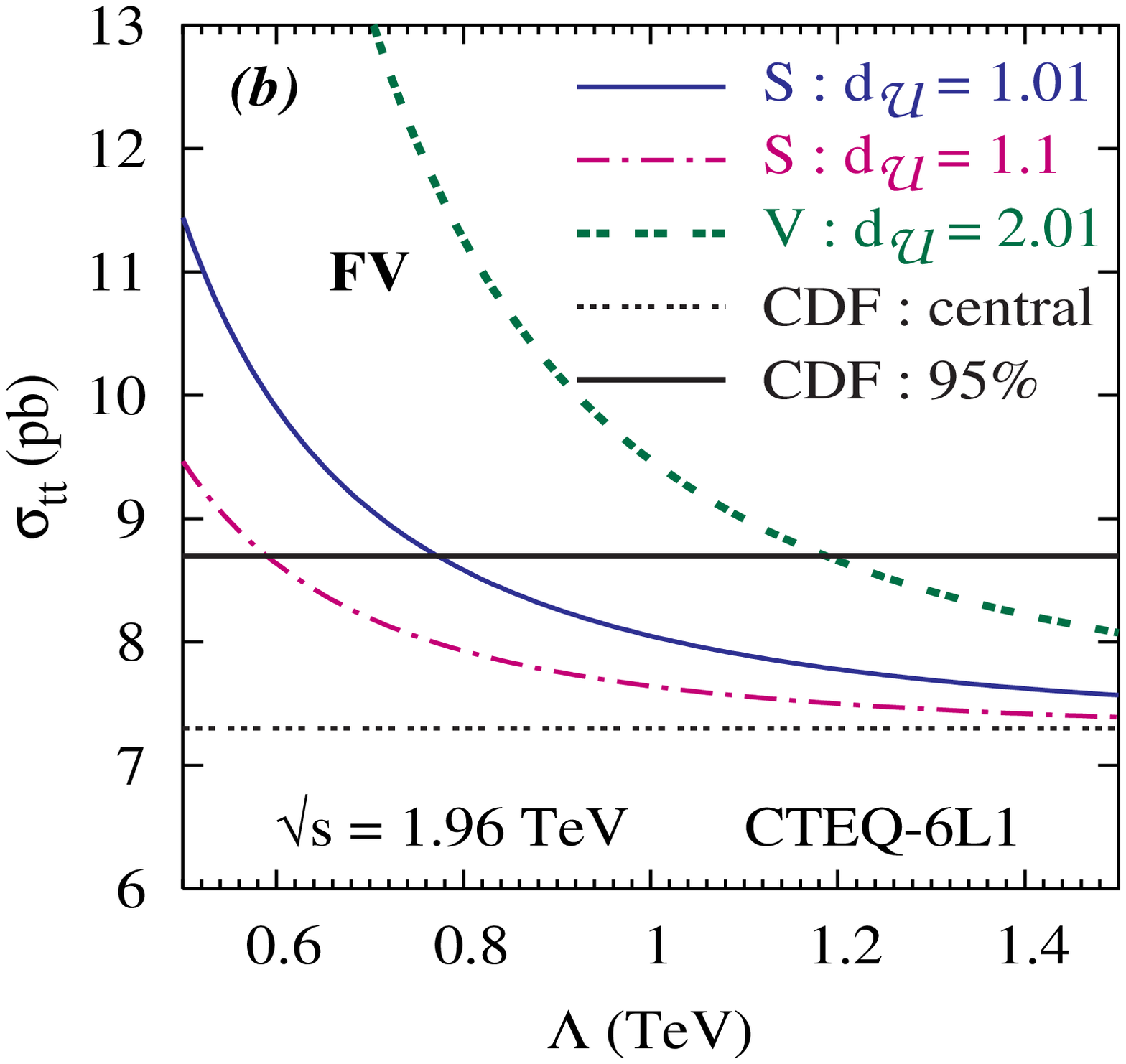}
\vspace*{0.5cm}
\caption{\em $\sigma(t \bar t)$ at Tevatron. {\em (a)}
flavour-conserving vector unparticle; 
{\em (b)} FV unparticle (both scalar and vector)
The horizontal lines correspond
to the current central value from the CDF experiment~\cite{Cabrera:2006ya}
and the $95 \%$ C.L. upper limit.}
\label{fig:cs_tev}
\end{center}
\end{figure}

In Fig.\ref{fig:cs_tev}$a$, we display the $t \bar t$ cross section 
at the Tevatron as a function of scale $\Lambda_{\cal U}$, in the presence
of a vector unparticle. With the latter being exchanged in the 
$s$-channel, there is no interference with the (dominant)  
QCD amplitude. Thus, unparticle effects can become appreciable only when 
the amplitude becomes comparable to the QCD one. 
With the electroweak contribution being very small, 
the famed phase factor 
$\exp(-i\pi d_{\cal U})$ in the unparticle propagator is
of little consequence.
As expected, for a given $d_{\cal U}$, the 
dependence on $\Lambda$ is power-law ($\Lambda^{4-4 d_{\cal U}}$). 
For comparison, we also display the current 
experimental data which gives (CDF Run II results averaged over 
all channels)~\cite{Cabrera:2006ya}
\[
\sigma(t \bar t)
 = 7.3 \pm 0.5 \, {(stat)} \pm 0.6 \, {(syst)} \pm 0.4 \, {(lum)} \ {\rm pb}\ .
\]
While the scalar unparticle amplitude is suppressed by 
the light quark mass, the tensor 
operator too does not give a substantial enhancement over the 
SM value. This can be easily understood in view of the stronger 
suppression in Eq.(\ref{lagr}). 

This being only an effective theory, flavour-violating (FV)
unparticle couplings  are also possible~\cite{Georgi:2007ek,fcnc_unp,Choudhury:2007js}. For example, the vector coupling can be generalized to 
\beq
{\cal L}_{FV} \supset
\Lambda^{1 -d_{\cal U}}\; 
\bar q \, \gamma_\eta \, (\tilde v^{qq'} + \tilde a^{qq'} \, \gamma_5) \, q' \; 
{\cal O}^\eta_V 
     \label{fcnclagr} 
\eeq 
and similarly for the scalar and tensor operators.  Note that
consistency demands that, for such FV couplings to be present, $d_{\cal U}
> 2$ for ${\cal O}_V$~\cite{Choudhury:2007js} and $d_{\cal U} > 3$ for
${\cal O}_T$. These operators result in a $t$-channel 
diagram for $u \bar u \to t \bar t$, which, of course, interferes with the 
QCD amplitude. 
In Fig.\ref{fig:cs_tev}($b$), we 
exhibit the corresponding $\sigma_{t \bar t}$
in the presence of either a scalar or a vector  FV type coupling. 

\begin{figure}[!h]
\begin{center}
\vspace*{-0cm}
\includegraphics[width= 8 cm, height= 5 cm]{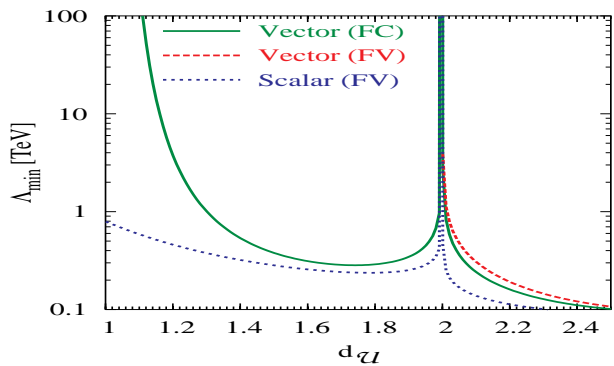}
\vspace*{-0.2cm}
\caption{\em The $95\%$ CL lower bound on the scale $\Lambda $ as a function
of $d_{\cal U}$ as 
obtained from $\sigma (t \bar t)$ at Tevatron Run II~\cite{Cabrera:2006ya}.
FC (FV) denotes flavour conserving(violating) 
operators.}
\label{fig:tev95cl}
\end{center}
\end{figure}

Using the aforementioned experimental determination of 
$\sigma_{t \bar t}$ (with errors added in quadrature), 
we may impose bounds on the unparticle parameter space, 
for a given choice of operators. In Fig.\ref{fig:tev95cl}, 
we display the 95\% C.L. bounds assuming that only one 
of the operators, whether flavour-conserving (FC) or FV, contributes.
As expected, the constraint 
is strongly dependent on the structure of the operator involved as well as on
the scale dimension $d_{\cal U}$.  The sharp rise of the
limit at $d_{\cal U} = 2 $ is but a manifestation of the presence of 
physical poles in the unparticle propagator at all integral values 
of $d_{\cal U} > 1$. It should be noted that these bounds are what can be 
achieved from the use of currently published total cross section data alone. 
Once more data is analysed, the constraints would only be stronger.
%
%

\begin{figure}[!h]
\begin{center}
\vspace*{-1.5cm}
\includegraphics[width= 10 cm, height= 6 cm]{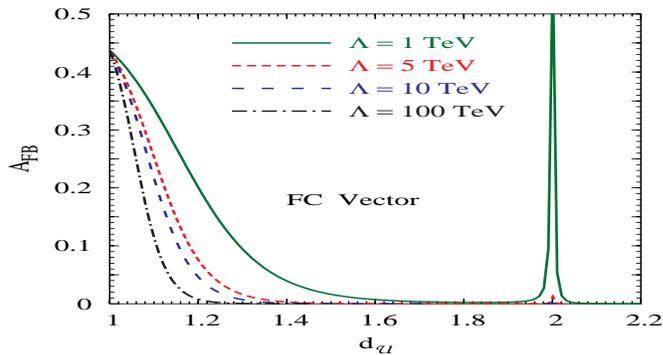}
\vspace*{0.2cm}
\caption{Forward-backward asymmetry in $t \bar t$ production caused by 
vector unparticle (right-handed FC coupling)}
\label{fig:FB}
\end{center}
\end{figure}

With unparticle operators being chiral in nature, they have the interesting 
consequence of leading to a potentially large forward-backward asymmetry 
(see Fig.\ref{fig:FB}). With $A_{FB}$ in the standard model being very 
small, this could potentially lead to added sensitivity. Furthermore, with 
the extent and the sign of the asymmetry being dependent on the nature of 
the coupling, this could also serve as a discriminator between scenarios.

Simultaneous presence of 
both FV and FC couplings could give rise to single-top production 
(viz. $u \bar u, d \bar d \to u \bar t$), processes which have no 
SM counterpart. The rates could be as large as $\sim 100 \, {\rm fb}$. 
While single-top production has been measured 
at the Tevatron~\cite{Abazov:2006gd}, the search strategy so far 
has explicitly assumed an associated $b$ in the hard process and,  
consequently, the said data cannot be readily used in this context.

\begin{figure}[!h]
\begin{center}
\vspace*{1.cm}
\includegraphics[width= 9 cm, height= 6 cm]{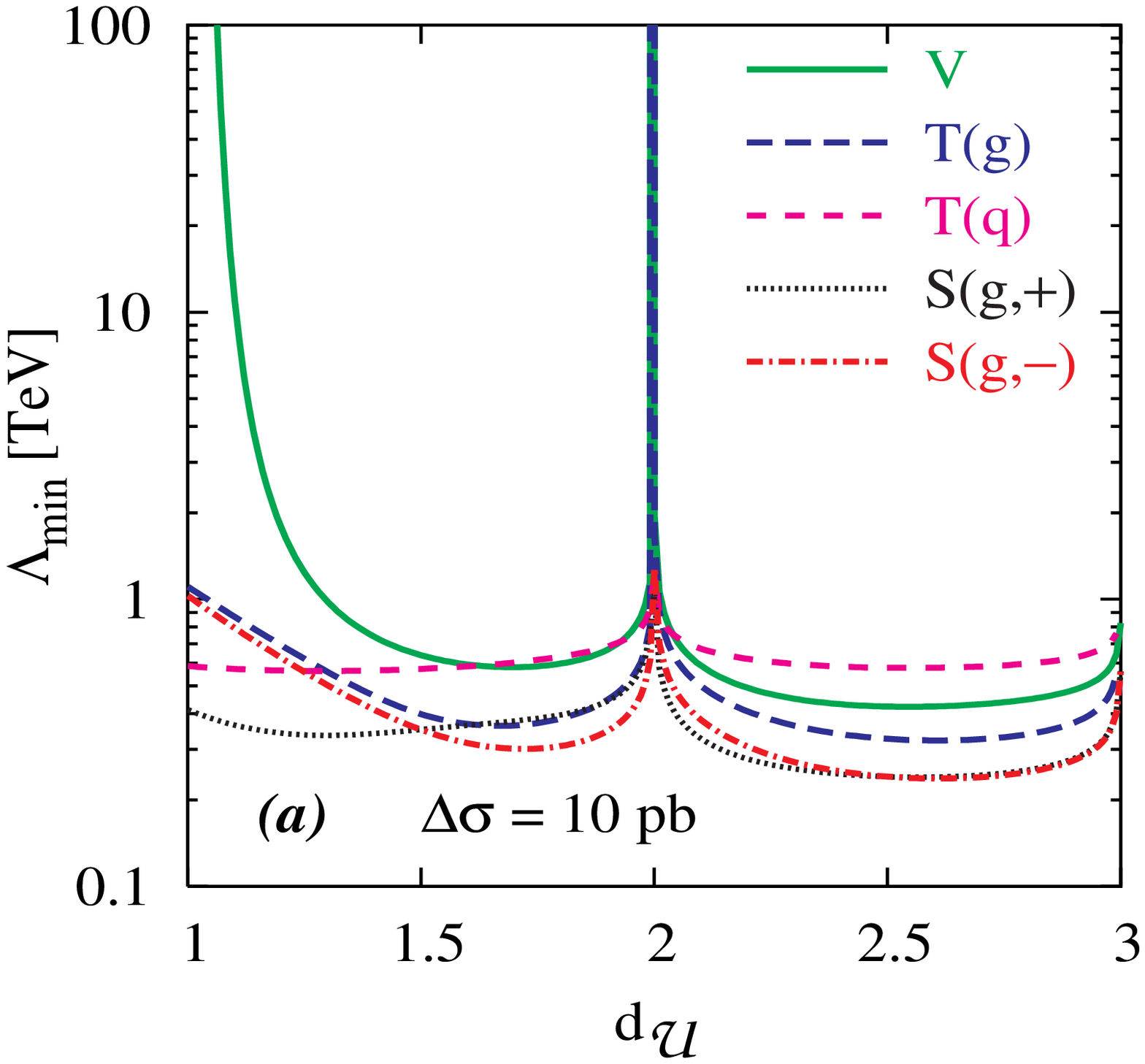}
\vspace*{-3.2cm}
\includegraphics[width= 9 cm, height= 6 cm]{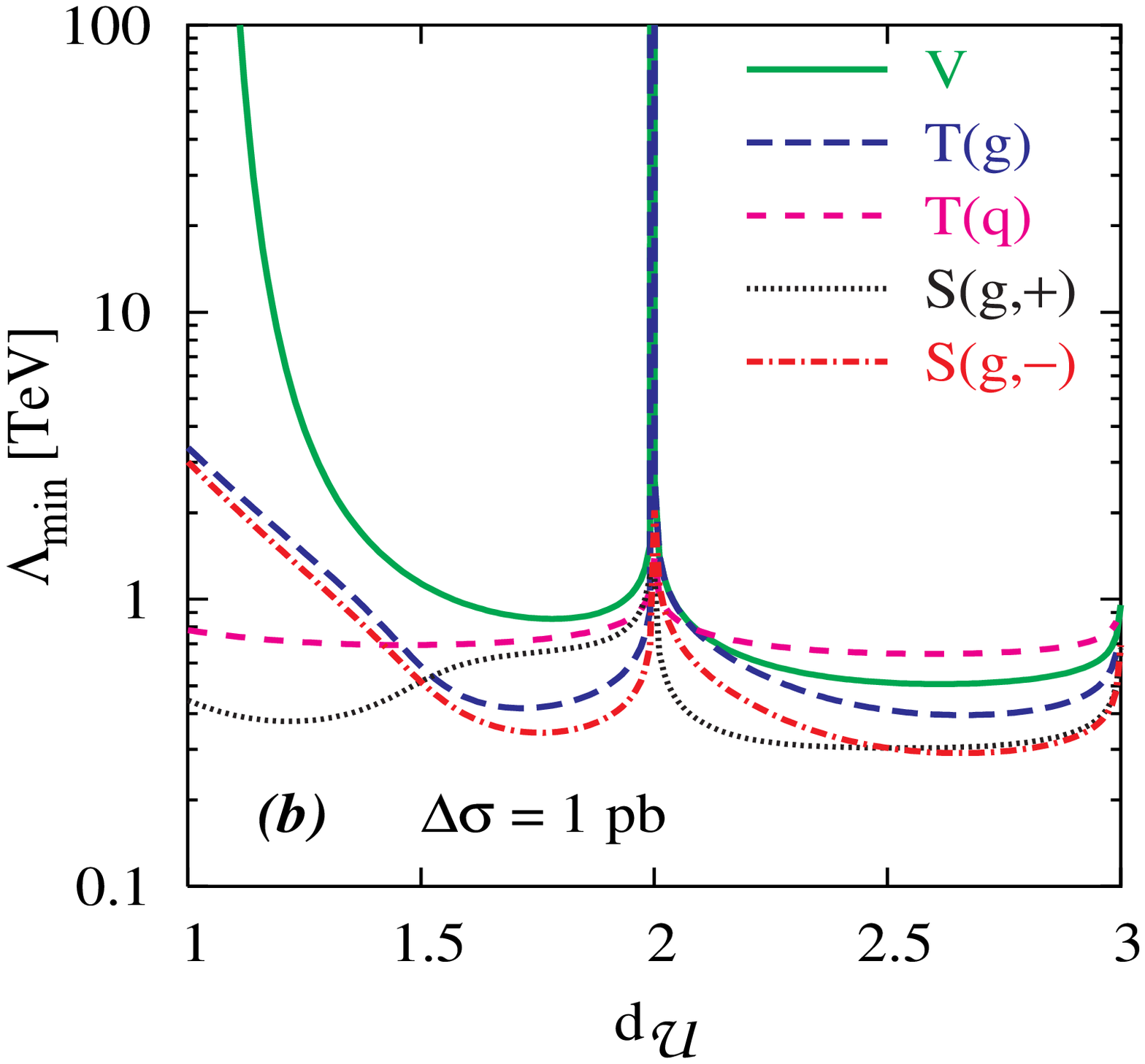}
\vspace*{0.8cm}
\caption{$3\sigma $ reach in $\Lambda $ as a function 
of $d_{\cal U}$ for flavour conserving operators
using the $\sigma(t \bar t)$  at the LHC.
$S(g, \pm)$ corresponds to the $gg \to {\cal O}_S \to t \bar t$ 
process with opposing signs of the product of gluon and top couplings.
Panel {\em (a)} assumes a measurement error of 10 pb and {\em (b)} of 1 pb.}
\label{fig:3sig_lhc}
\end{center}
\end{figure}

We now turn our sights on the forthcoming LHC. The situation here is 
somewhat different. With the much larger gluon flux, the r\^ole of 
unparticle mediation in gluon-initiated $t\bar t$ production 
assumes importance and we begin to be able to probe such couplings. 
However, since the gluons couple only to ${\cal O}_S$ and ${\cal O}_T$, 
the constraints are always weaker than those derivable for 
the ${\cal O}_V$ mediated process initiated from $q \bar q$. In anticipating 
the bounds, we make an assumption that the $t \bar t$ cross section 
($\sim 830$ pb in the SM) would be determined to an accuracy of 10 pb 
(a conservative estimate given the accuracy at Tevatron and the expected 
luminosity at the LHC). The corresponding $3 \sigma$ reaches are displayed 
in Fig.\ref{fig:3sig_lhc}$a$. Note that, with the relative signs of the 
unparticles couplings with the top and the gluon being undetermined, the 
interference with the QCD diagrams may have either sign. The consequent 
effect 
has been illustrated for ${\cal O}_S$ in Fig.\ref{fig:3sig_lhc}. 
At large $d_{\cal U}$, away from integral values, the reach in $\Lambda$ 
tends to have very little dependence on $d_{\cal U}$. This is easily
 understood by realizing that the unparticle amplitude typically behaves 
as $(\hat s / \Lambda^2)^\gamma$, where $\gamma = d_{\cal U}, \, 1-d_{\cal U}$.
With the bounds already in the regime 
of $ \Lambda \sim  2 \, m_t$, 
changing $d_{\cal U}$ naturally has very little effect. Improvement in the 
$\sigma_{t \bar t}$ measurement does have an effect, though (see 
Fig.\ref{fig:3sig_lhc}$b$). The improvement is marginal at larger 
$d_{\cal U}$ on account of the power law suppression.
It should be noted here that further improvements in the sensitivity 
may be possible if one were to consider phase-space distributions of 
the $t \bar t$ pair~\cite{axigluon}
 or the spin-spin correlations~\cite{Zpr,Arai:2004yd}

\begin{figure}[!h]
\begin{center}
\vspace*{-1.0cm}
\includegraphics[width= 10 cm, height= 6 cm]{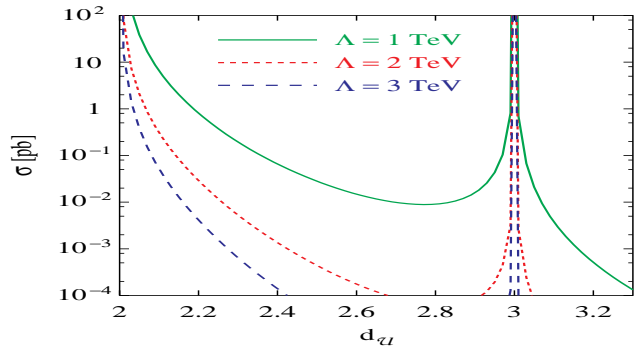}
\vspace*{-0.2cm}
\caption{Cross section for $t t$ (like sign) production at the LHC through FV 
vector unparticle.}
\label{fig:like_sign}
\end{center}
\end{figure}

While all our discussion so far about the LHC  has been confined to 
FC coupling, FV couplings can be probed similarly, both 
through $t\bar t$ as well as single-top  production. It 
is amusing to consider though the possibility of producing 
like-sign tops (through $u u \to t t$), particularly because it is 
precluded at the Tevatron on account of the low $u u $ flux. As
Fig.\ref{fig:like_sign} testifies, the rates at the LHC could be sizable, 
leading to spectacular signals. 


In summary, we have probed the effect of unparticle-couplings of the
SM fields (matter and gauge) in top pair production at the Tevatron
and the LHC. The current Tevatron measurements can be used to impose
significant constraints on unparticle physics.  The bounds are
strongly dependent on the Lorentz structure of the relevant unparticle
operator as well as its mass dimension $d_{\cal U}$. For
flavour-conserving vector couplings, the bound on the scale $\Lambda$
could be as large as several hundred TeVs for $d_{\cal U}$ close to
1. For larger $d_{\cal U}$, the bounds get progressively weaker by a
power-law since the amplitude scales as $\Lambda^{2-2 d_{\cal U}}$.
With the theory also allowing flavour-violating interactions,
additional contributions to $u \bar u \to t \bar t $ may accrue.  As
such amplitudes interfere with the QCD one (unlike the FC ones), the
bounds are slightly stronger for identical $d_{\cal U}$ (it should be
remembered though that the FV operators are restricted to have higher
$d_{\cal U}$). As for the scalar and tensor operators, the 
constraints are understandably weaker. 
Analysis of further data by CDF/D0 would not only 
strengthen these bounds, but also allow for the exploitation of 
forward-backward asymmetry, which, in such models, can be very large 
indeed.

At the LHC, if one assumes an experimental accuracy in $\sigma_{t \bar t}$ 
of only 10 pb, the bound on FC vector unparticle 
improves only marginally for smaller $d_{\cal U}$. For larger $d_{\cal U}$ 
the improvement is significant compared to Tevatron bound. 
An accuracy at the 1 pb 
level, changes the situation considerably. In addition, large luminosities 
would also allow the exploitation of the distortion in the $t\bar t$ 
spectrum, thus increasing sensitivity. More importantly, the LHC 
allows us to probe the gluonic couplings of the unparticles in the 
$t \bar t$ mode, which the Tevatron is insensitive to. 
Finally, the presence of FV violating couplings 
could lead to like sign top pair production. Even with parameters 
perfectly in agreement with Tevatron data, the production rate can be as high
as 100 pb, leading to spectacular signatures. 
We thus hope that the analysis presented here would  encourage 
our colleagues at both the Tevatron and the LHC to carry out 
more detailed investigations to probe
the curious world of unparticles.

DC thanks  the 
DST, India for  support under project 
SR/S2/RFHEP-05/2006. DKG thanks the Theory
Division, CERN and the HECAP Section of the AS-ICTP for hospitality 
while part of the work was completed.


\end{document}